\documentclass[reprint,amsmath,amssymb,aps]{revtex4-1}
\usepackage{graphicx}
\usepackage{dcolumn}
\usepackage{bm}

\usepackage{color}

\usepackage{hyperref}
\hypersetup{ colorlinks=true, linkcolor=blue, citecolor=red, urlcolor=blue }

\usepackage{physics}

\begin{document}

\title{Experimental Machine Learning of Quantum States}

\author{Jun Gao$^{1,2}$, Lu-Feng Qiao$^{1,2}$, Zhi-Qiang Jiao$^{1,2}$, Yue-Chi Ma$^{3,4}$, Cheng-Qiu Hu$^{1,2}$, Ruo-Jing Ren$^{1,2}$, Ai-Lin Yang$^{1,2}$, Hao Tang$^{1,2}$, Man-Hong Yung$^{4,5*}$}
\author{Xian-Min Jin$^{1,2}$}
\email{yung@sustc.edu.cn}
\email{xianmin.jin@sjtu.edu.cn}

\affiliation{$^1$State Key Laboratory of Advanced Optical Communication Systems and Networks, Department of Physics and Astronomy, Shanghai Jiao Tong University, Shanghai 200240, China}
\affiliation{$^2$Synergetic Innovation Center of Quantum Information and Quantum Physics, University of Science and Technology of China, Hefei, Anhui 230026, China}
\affiliation{$^3$Center for Quantum Information, Institute for Interdisciplinary Information Sciences, Tsinghua University, Beijing 100084, China}
\affiliation{$^4$Institute for Quantum Science and Engineering and Department of Physics, Southern University of Science and Technology, Shenzhen 518055, China}
\affiliation{$^5$Shenzhen Key Laboratory of Quantum Science and Engineering, Shenzhen, 518055, China}

\begin{abstract}
Quantum information technologies provide promising applications in communication and computation, while machine learning has become a powerful technique for extracting meaningful structures in `big data'. A crossover between quantum information and machine learning represents a new interdisciplinary area stimulating progresses in both fields. Traditionally, a quantum state is characterized by quantum state tomography, which is a resource-consuming process when scaled up. Here we experimentally demonstrate a machine-learning approach to construct a quantum-state classifier for identifying the separability of quantum states. We show that it is possible to experimentally train an artificial neural network to efficiently learn and classify quantum states, without the need of obtaining the full information of the states. We also show how adding a hidden layer of neurons to the neural network can significantly boost the performance of the state classifier. These results shed new light on how classification of quantum states can be achieved with limited resources, and represent a step towards machine-learning-based applications in quantum information processing.
\end{abstract}

\maketitle

Over the last few years, there has been a significant advancement in an emerging field of quantum machine learning~\cite{Seth2017,Dunjko2017}, where quantum information meets the modern information-processing technologies. On one hand, various modern quantum technologies, such as quantum communication~\cite{BB84,QKDrev,Pan}, quantum computation~\cite{Shor,Grover} and quantum metrology have been experimentally demonstrated by exploiting quantum entanglement as a resource~\cite{Jin2010,Lu2007}. On the other hand, machine learning, a modern technique for making predictions by mining information from `big data'~\cite{ML}, has been proven as one of the most successful achievements in artificial intelligence. Notable examples include self-driving cars and the famous Alpha-Go which surpasses the top human players at the game `\emph{Go}'~\cite{Go,Go0}. 

The key question in quantum machine learning is, how to develop new ideas for applying technologies in machine learning to quantum information, or vice versa, to gain advancements in both fields? In fact, several promising steps along both directions have already been taken in the community. In particular, machine learning can in principle exploit quantum superposition to enhance its performance. For example, quantum versions of the principal component analysis (PCA)~\cite{PCA} and support vector machines (SVM)~\cite{SVM} have been proposed. Other quantum algorithms along with their experimental implementations~\cite{Seth,Krenn2016,Melnikov2018,Cai2015,Li2015} have also been demonstrated in recent years to broaden the versatility of machine learning. 

Besides, machine learning can also be applied to certain quantum tasks, from classifying separability of quantum states~\cite{Yung2017,Lu2017} to classifying phases in condensed matter physics~\cite{Phase1,Phase2}, and even the development of new classical algorithms for solving many-body systems~\cite{MBS}. These results suggest that machine learning of quantum states represents a new platform for solving problems in quantum information science. 

Here we report an experimental machine learning of quantum states, where an artificial neural networks (ANN) is trained for classifying the separability of some quantum states, given that only {\it partial information} about the quantum states is available. More specifically, based on the experimental data, we have constructed a quantum-state classifier~\cite{Yung2017}, which generalizes the pattern recognition in learning theory for quantum data. In the classical setting, a set $\{x_i,y_i\}$ of data $x_i$ and label $y_i$ are supplied as the training set for the machine-learning program, and the output is a classifier (a function or a program) for predicting labels of new data. In the quantum setting, the data $x_i$ may be replaced with a density matrix $\rho_i$, and the corresponding label may be taken as any physical property, e.g. separability. However, the size of a quantum state grows exponentially when scaled up, which makes large-scale quantum state tomography~\cite{Tomo,Entrev} intractable to carry out. Meanwhile, the task of quantum state classification does not require complete information of the density matrix. This motivates us to exploit the possibility of learning with only partial information of the quantum state.

\begin{figure}
 \centering 
 \includegraphics[width=0.98\columnwidth]{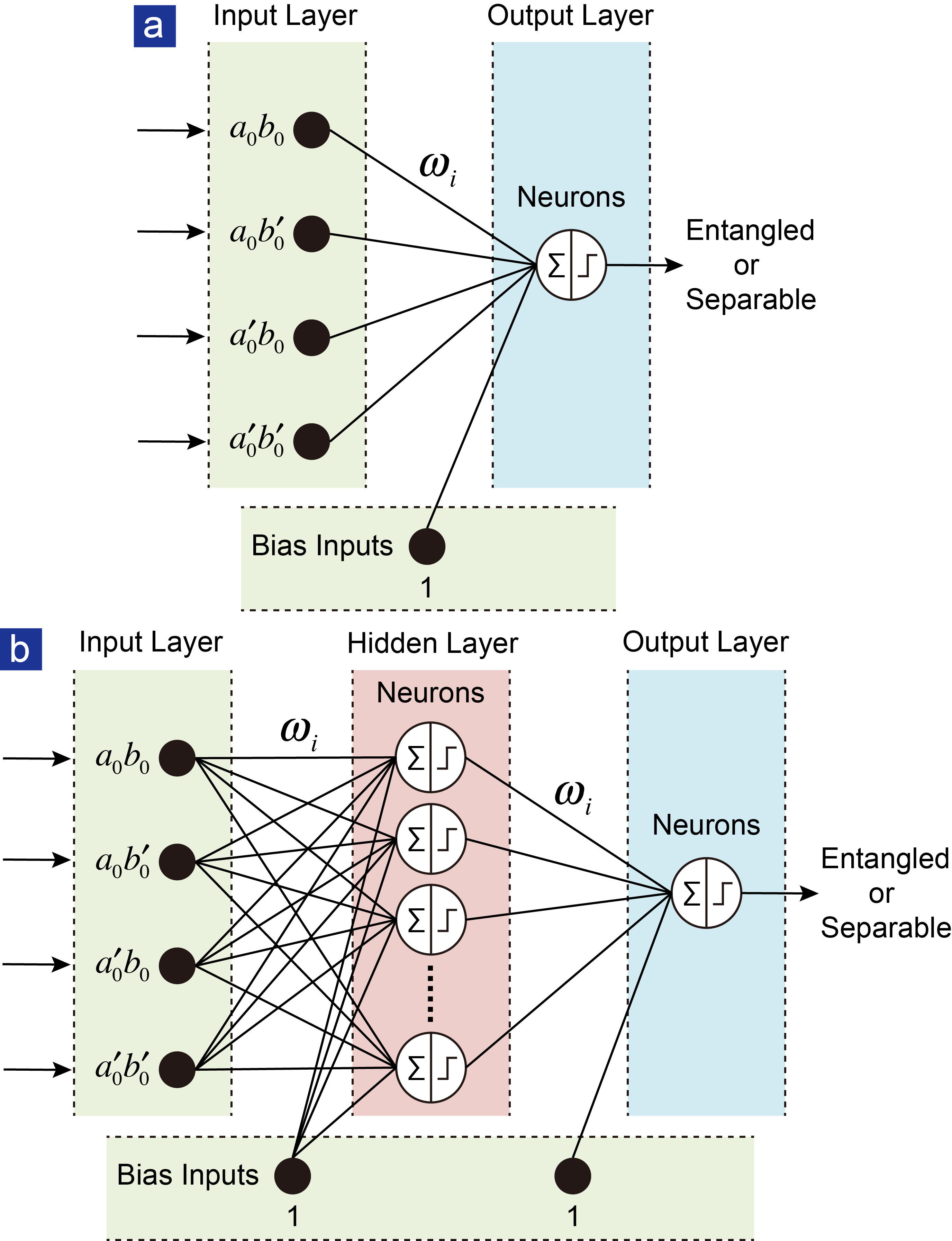}\\
 \caption{{\bf ANN-based quantum state classifier.} (a) Linear ANN optimizes new weight coefficients $\{\omega_i\}$ with 4 observables $\{ \left\langle {{a_0}{b_0}} \right\rangle ,\left\langle {{a_0}b_0^\prime } \right\rangle ,\left\langle {a_0^\prime {b_0}} \right\rangle ,\left\langle {a_0^\prime b_0^\prime } \right\rangle \}$ as input features. (b) ANN with hidden layers. To further promote the performance of the classifier, we insert an additional hidden layer, where each neuron conducts a non-linear function on the input features.}
 \label{FIG. 1.}
\end{figure}

For the purpose of illustration, we consider a toy model (see Supplementary Material A), which take a class of Werner-like states~\cite{Werner} as the training and testing sets. The label on the separability is determined by using the positive partial transpose (PPT) criteria~\cite{PPT,Qutip}, applied only to the training set, but not the testing set. Experimentally, we first show that a linear optimization of the Clauser-Horne-Shimony-Holt (CHSH) inequality~\cite{Bell,CHSH} can significantly boost the accuracy of the optimized CHSH inequality in identifying the separability of the quantum state of a pair of qubits. Although it is still far from being perfect, the results in the testing phase exceed the conventional CHSH inequality in detecting entanglement. In the second part, we further equip the neural network with a hidden layer of neurons, making it a non-linear model. We compare the performance of the two neural networks and find that the inclusion of a hidden layer can significantly improve the match rate of the classifier to nearly unity. The experimental result is consistent with the speculation~\cite{Yung2017} that the inclusion of the hidden layer can be regarded as an optimization of multiple linear CHSH-like inequalities. Finally, we show that the performance decreases, if the classifier is instead trained with theoretical values, but tested with experimental values, suggesting that experimental training of the neural network is necessary for the construction of the classifier for testing experimental data. 

A violation of the CHSH inequality implies that the quantum state is entangled. However, the opposite is not true; there exists entangled quantum states which does not violate the CHSH inequality (see Supplementary Material B).  We can construct a linear state classifier by the following CHSH operator $\Pi_{ml}$, 
\begin{equation}
\Pi_{ml}\equiv w_1a_0b_0+w_2a_0b^\prime_0+w_3a^\prime_0b_0+w_4a^\prime_0b^\prime_0+w_0 \ ,
\end{equation}
which contains a set of parameters $\{w_0, w_1, w_2, w_3, w_4\}$ to be optimized in a state-independent way~\cite{Yung2017}. For the original CHSH inequality, the unoptimized values are $\{1,-1,1,1,-2\}$. Here, the measurements are given by $a_0=\sigma_z$, $a^\prime_0=\sigma_x$, $b_0=(\sigma_z+\sigma_x)/\sqrt{2}$ and $b^\prime_0=(\sigma_z-\sigma_x)/\sqrt{2}$. Moreover, as shown in Fig.1(a), each time we take as an input to the ANN only 4 observables (or `features' in the language of machine learning) $\{ \left\langle {{a_0}{b_0}} \right\rangle ,\left\langle {{a_0}b_0^\prime } \right\rangle ,\left\langle {a_0^\prime {b_0}} \right\rangle ,\left\langle {a_0^\prime b_0^\prime } \right\rangle \}$, instead of 16 observables as in quantum state tomography~\cite{Tomo}. 

To further improve the performance of the classifier, we connect the input layer to a hidden layer, as shown in Fig.1(b). In addition, the measurements are no longer restricted in the $x$-$z$ plane. The input vector is still given by the 4 observables, denoted as $\vec{x}_0$. Next, an intermediate vector $\vec{x}_1$ is constructed through a non-linear relation for every neuron in the hidden layers,
\begin{equation}
\vec{x}_1=\sigma_{RL}(W_1\vec{x}_0+\vec{w}_{01}) \ ,
\end{equation}
where $\sigma_{RL}$ is the ReLu function. The matrix $W_1$ and the vector $\vec{w}_{01}$ are initialized uniformly and optimized through the learning process. Then the net step is to optimize the following function,
\begin{equation}
{\sigma _S}({W_2}{{\vec x}_1} + {{\vec w}_{02}}) \ ,
\end{equation}
where $\sigma_{S}=1/(1+e^{-x})$ is the sigmoid function. Here the number of neurons (denoted as $n_{ne}$) in the hidden layer can be varied to obtain the optimal performance. 

\begin{figure*}
 \centering
 \includegraphics[width=1.9\columnwidth]{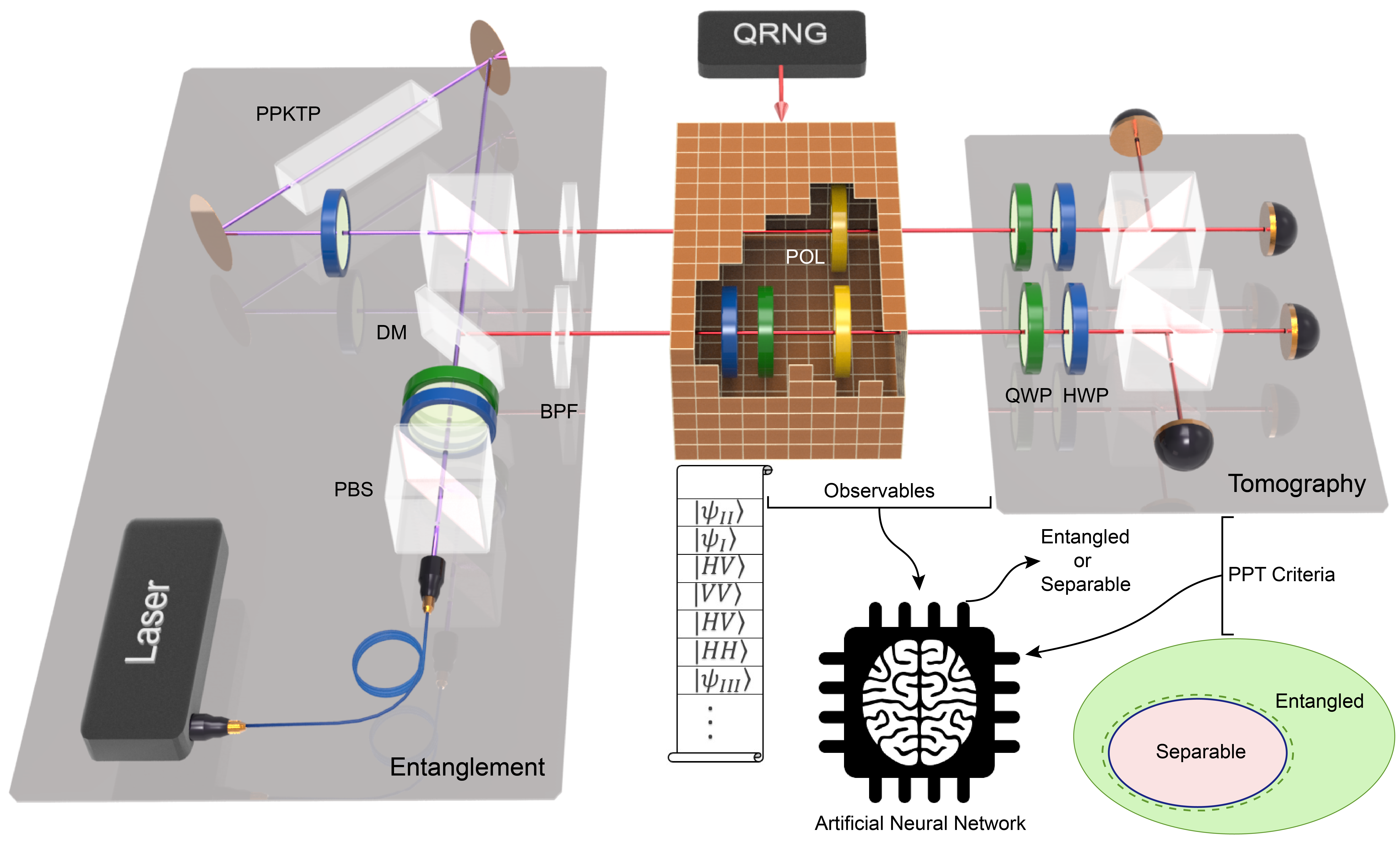}\\
 \caption{\textbf{Experimental setup of the quantum state classifier.} The prepared entanglement is guided to the state generator (yellow box). Either relative phases can be added between the two components or the state can be projected to four components $\arrowvert{H_AH_B}\rangle$, $\arrowvert{H_AV_B}\rangle$, $\arrowvert{V_AH_B}\rangle$ and $\arrowvert{V_AV_B}\rangle$ by the polarizers (POL) to construct the identity state. All the generated states are then analyzed by both the quantum state tomography and the projection measurements. Theoretically predicted labels made by the PPT criteria based on the tomographic data are also added to the desired states. A quantum random number generator (QRNG) picks different components to construct mixed states as the training or the testing states. All data are sent to the agent for the training and the testing stages.}
 \label{FIG. 2.}
\end{figure*}

In the following, we report the experimental results in constructing the quantum-state classifier, where the experimental setup is shown in Fig.2. Polarization-entangled photon pairs are created through a type-II spontaneous parametric down conversion in a quasi-phase matched periodically-poled KTiOPO$_4$ (PPKTP) crystal based on the Sagnac interferometer~\cite{PPKTP}. The $405$-nm-pump laser is first coupled into a single mode fiber to acquire a near-perfect transverse mode, and is prepared as a superposition state, $\cos\theta\arrowvert{H}\rangle+e^{i\phi}\sin\theta\arrowvert{V}\rangle$, by combining a half-wave plate (HWP) with a quarter-wave plate (QWP). The pump light is then divided into two directions and focused on the crystal. Through a careful alignment of the Sagnac interferometer, the clockwise and anti-clockwise components become indistinguishable, generating the following entangled state,
\begin{equation}
\arrowvert{\psi_{AB}}\rangle=\cos\theta\arrowvert{H_AV_B}\rangle+e^{i\phi} \sin\theta\arrowvert{V_AH_B}\rangle \ .
\end{equation}
The photons are then coupled into two single-mode fibers and are spectrally filtered by two bandpass filters (BPF). By adjusting the HWP and QWP, we can control the parameter $\theta$ and $\phi$ with a high precision to generate a family of entangled states. We characterize the quality of the entangled state by quantum state tomography. The concurrence is found to be $0.927$ and the purity is $0.914$.

The next step is to construct a series of desired quantum states which will be used for both the training and the testing stage. In our work, we use the time-mixing technique~\cite{Borivoje2012,Dylan2017} to create the Werner-like states,
\begin{equation}\label{WLS}
\rho_W=p\arrowvert{\psi_{AB}}\rangle\langle{\psi_{AB}}\arrowvert+(1-p)I/4  \ ,
\end{equation}
for $0\le p\le1$. Here $\arrowvert{\psi_{AB}}\rangle\langle{\psi_{AB}}\arrowvert$ is the entangled state generated by the PPKTP source, and the identity matrix is obtained by collapsing the wave function to the following four components, $\arrowvert{H_AH_B}\rangle$, $\arrowvert{H_AV_B}\rangle$, $\arrowvert{V_AH_B}\rangle$ and $\arrowvert{V_AV_B}\rangle$. The parameters $\theta$ and $\phi$ are manipulated by the rotation of the HWP and QWP. We conduct both the quantum-state tomography and the CHSH measurements for all the component states. All these data have been saved as a data pool. In order to eliminate potential time-dependent fluctuation of the source, we randomly picked the corresponding components from the pool to construct the desired density matrix of the Werner-like states, and the four observables under the projection. We employed a quantum random number generator to control random selection to avoid potential artificially introduced bias.

\begin{figure}
 \centering
 \includegraphics[width=0.8\columnwidth]{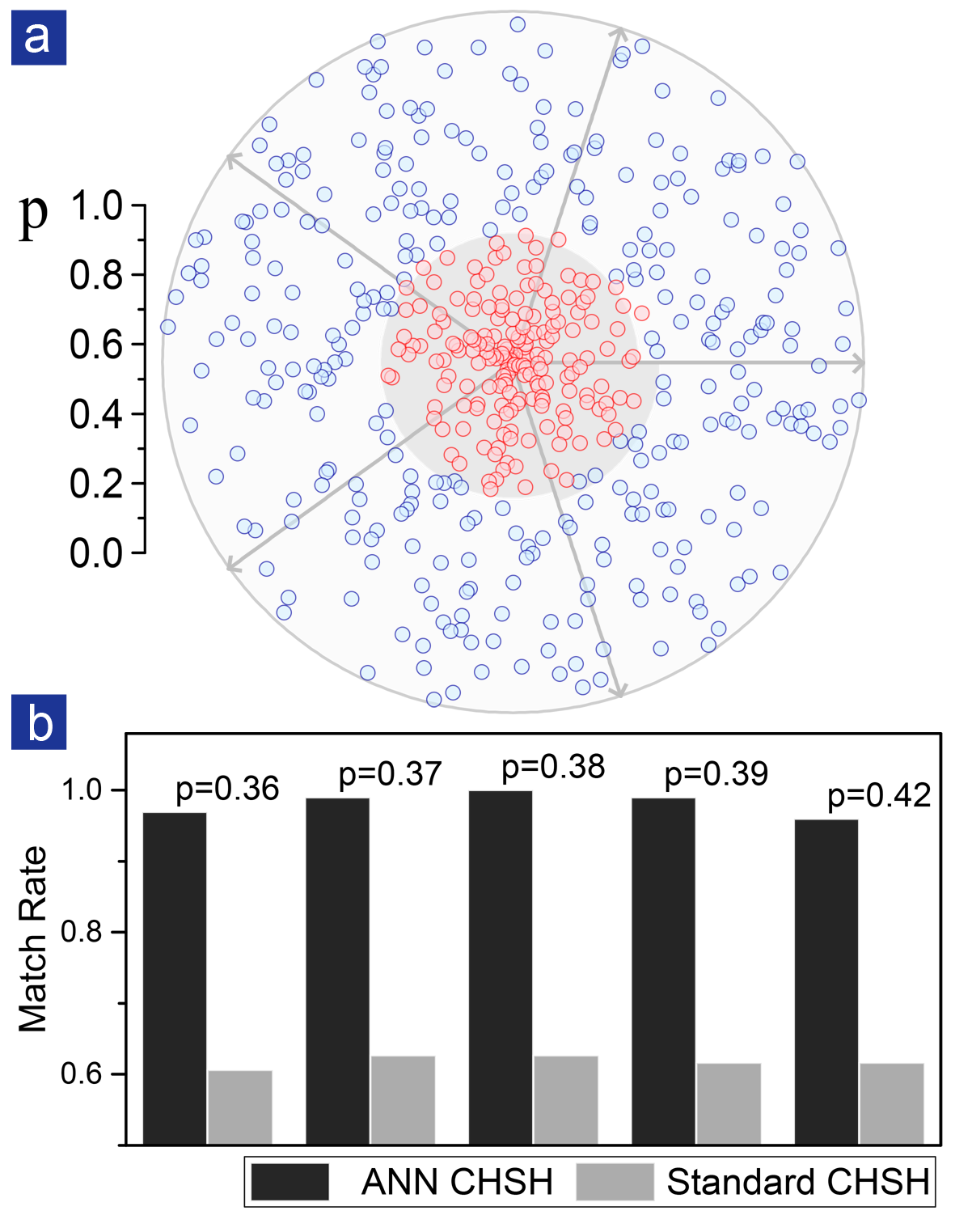}\\
 \caption{
  \textbf{Results predicted by the linear ANN.} (a) Five sections correspond to five different parameters $\theta$ which set the separable-entangled bound $p$ predicted by the PPT criteria (the light gray part marks the separable section). All the data points are labeled by the linear ANN, here blue represents the entangled label while red represents the separable label. (b) Comparison of match rate between standard CHSH inequality and liner ANN classifier. The performance of optimized weight coefficients significantly surpass the standard CHSH coefficients.}
 \label{FIG. 3.}
\end{figure}

We first study the performance of the linear CHSH classifier. We fix the relative phase between the two components of the entangled state and vary the parameter $\theta$ with 5 different values. For each value of $\theta$, we prepare 99 Werner-like states with uniform $p$ distribution from 0.01 to 0.99. These states are used as the training set, and the labels (entangled or separable) are determined by the PPT criteria on the experimental density matrices. After training the ANN, we obtain an optimized weight coefficients $\{30.54, -32.42, -1.219, -0.3819, 15.62\}$. 

\begin{figure}
 \centering
 \includegraphics[width=0.98\columnwidth]{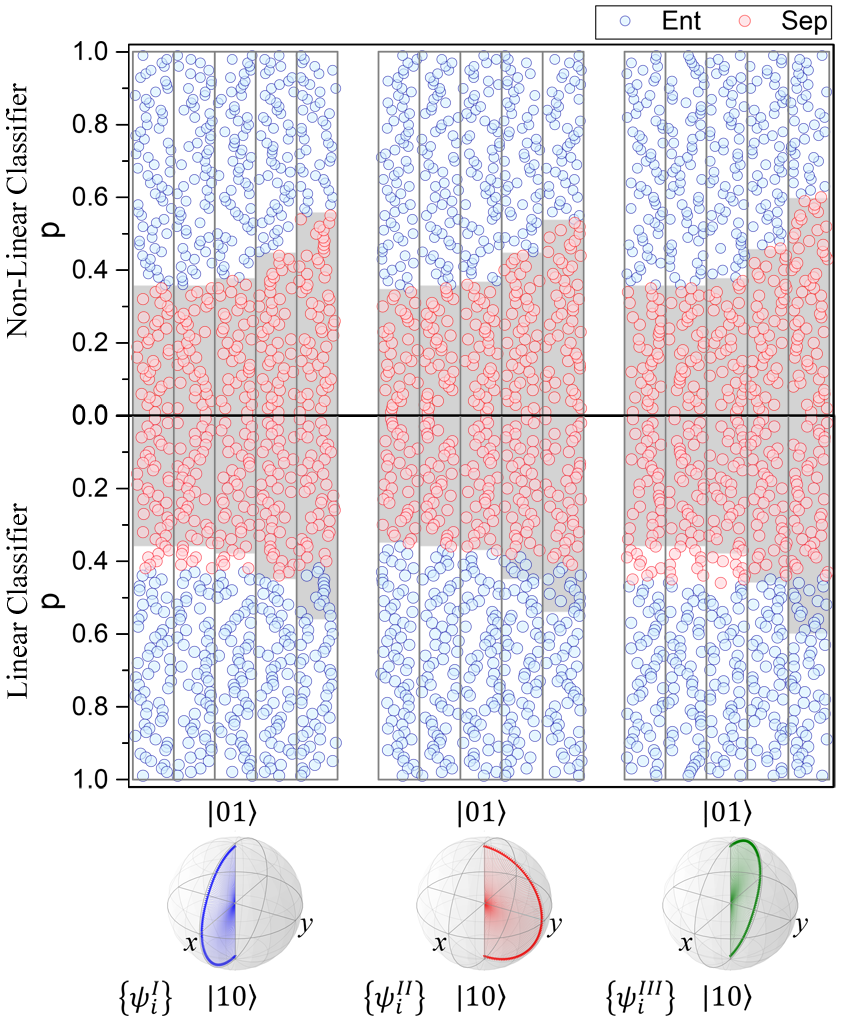}\\
 \caption{
  \textbf{Performance comparison between non-linear and linear classifiers.} Three classes of states $\{\psi^{I}_i\}$, $\{\psi^{II}_i\}$ and $\{\psi^{III}_i\}$ are prepared to compare the results between the non-linear and linear ANN. The gray shade gives the PPT predicted separable region. The ANN predicted labels are marked with different colors. Non-linear ANN can predict the exact margin for each kind of state while the linear ANN shows its insufficiency. The Bloch spheres at the bottom mark the tunable parameter in the three classes of states.}
 \label{FIG. 4.}
\end{figure}

\begin{figure}
 \centering
 \includegraphics[width=0.98\columnwidth]{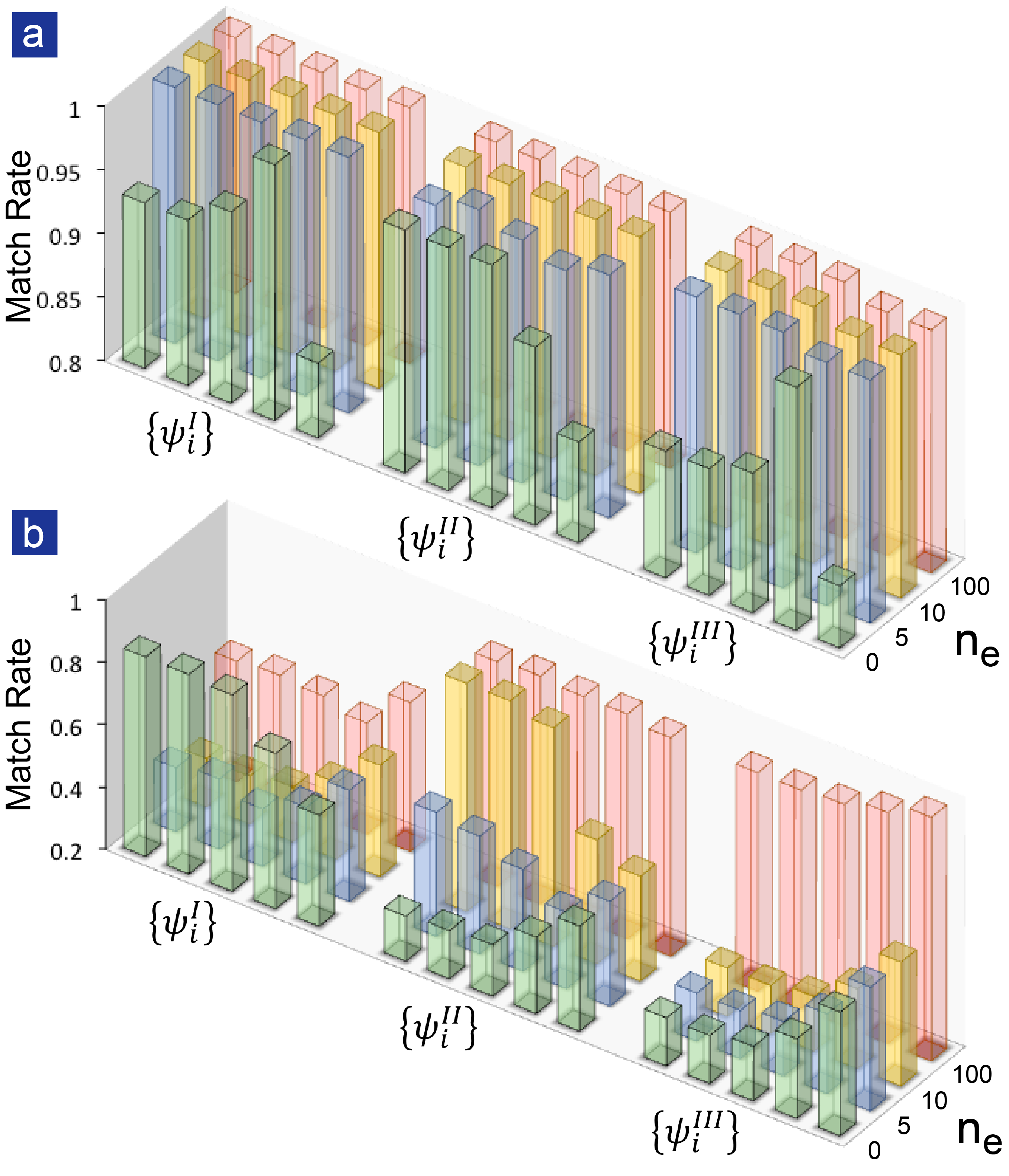}\\
 \caption{
  \textbf{Comparison of match rates with the classifier trained by theoretical data.} If the classifier is trained with theoretical data instead, the performance (b), in terms of the match rate, of it in predicting the experimental data is not as good as the one trained experimental data (a). In particular, the match rate of several instances in the class $\{\psi^{I}_i\}$ cannot be improved by increasing the number of hidden neurons.}
 \label{FIG. 5.}
\end{figure}

With these new weight coefficients we test the performance of the classifier and present the results in Fig.3(a). We divide the whole pie into 5 sections to correspond to different $\theta$ values. The classification results are compared with the theoretical prediction made by the PPT criteria. Most states are classified with accurate labels and the total match rate of the result is $98.3\%$. Another feature is that mismatch mainly occurs near the margin. This is because the weight coefficients are very sensitive near the bound. It should also be noticed that classifier exceeds the standard CHSH bound of $1/\sqrt{2}$ and can be applied to a wider range of states. We also compare the performance of these two classifiers in Fig.3(b). It is noteworthy that due to the uniform distribution of $p$ in our test states, the standard CHSH can still predict a bound to label the rest states above the bound as entangled leading to a considerable match rate, but the performance of the CHSH classifier cannot be considered as neither reliable or stable (see Supplementary Material C).

Next, we proceed to the case of ANN with a hidden layer. Unlike the linear classifier, we make several modifications in the training stage. We first prepare three different classes of states with different relative phases between the two components of the entangled state $\arrowvert{\psi_{AB}}\rangle$. By inserting an additional HWP or QWP, we can change the relative phase from 0 to $\pi/2$ or $\pi$, as plotted in the Bloch spheres in Fig.4. We denote these classes as $\{\psi^{I}_i\}$, $\{\psi^{II}_i\}$ and $\{\psi^{III}_i\}$. In each class, we also vary the parameter $\theta$ with five different values so we have prepared totally 15 different $\arrowvert{\psi_{AB}}\rangle\langle{\psi_{AB}}\arrowvert$ states. Besides, the measurement settings have also been changed to the combination of $\sigma_x$, $\sigma_y$ and $\sigma_z$ since additional information of the phase should be acquired. Last but not least, we deliberately pick states near the entangled-separable boundary as the training set to improve the learning process. For each Werner-like state, we select 80 states with $p\pm0.05$ near the margin. The testing states, however, still have the uniform distribution of $p$ from 0.01 to 0.99.

To further investigate the performance of the non-linear ANN, we vary the number of neurons $n_{e}$ in the hidden layers, from 0 to 5, 10 and 100. The experimental results shown in Fig.4 clearly indicate that the inclusion of the hidden layer can improve the performance, especially when we look at the most fallible and sensitive margin part. The average match rate increases from $93.3\%$ to $99.7\%$ (see Fig.5(a)). From these results we can also see that the linear classifier tends to predict an average bound for each class of state $\{\psi_i\}$ while the ANN with hidden layers can accurately predict the margin for every kind of state. The experimental results show that when dealing with more general scenario, ANN with a hidden layer (even with small numbers of neurons) can significantly enhance the performance of the classifier. 

Finally, we compare the performance of the classifier if theoretical data is trained instead of experimental data. As shown in Fig.5, the performance of the theoretical classifier is not as good as the one trained with experimental data. We observe a steady increase of match rate proportional to the number of the neurons $n_{e}$ (Fig.5(a)). In contrast, training with theoretical data only gives the classifier an ideal scenario and therefore may lead to errors when doing test with real experimental data. As is shown in Fig.5(b), the effect of the number of neurons $n_{e}$ is unclear and lack of apparent tendency, and the performance of classifier is state dependent. These results imply that the machine-learning program does take into account experimental noise. Consequently, it implies that theory-only investigations must be considered with caution. Our work present an experimental approach in applying machine learning for data analyzation, where we successfully demonstrated the applicability of machine learning in a realistic quantum scenario.

In summary, we experimentally demonstrate quantum machine learning of quantum entanglement by constructing a quantum-state classifier via artificial neural network. We show that a linear optimization of the neural network can already outperform the standard CHSH inequality, in terms of classifying quantum non-separability. The quantum-state classifier achieves an average match rate of $98.3\%$. We further demonstrate that ANN with a hidden layer can be applied to more general quantum states with an average match rate of $99.7\%$. Overall, the experimental results confirm the working principle of a quantum-state classifier in a small quantum system, where entanglement is taken as the label. 



\section*{Appendix}

\subsection{Remarks on the machine-learning method}
Quantum-state classifiers are generalization of classifiers for classical data in pattern recognition. At the conceptual level, it can be applied in the scenarios of a two-party game: suppose Alice and Bob are separated by a distance for performing an experiment (e.g., testing quantum nonlocality) utilizing a certain entangled state (not necessarily a Bell state), which is generated by a black-box machine far away from them. Suppose the machine is not so reliable, in the sense that sometimes it does generate the correct entangled state, but it may generates random bits. The task is to determine if, overall, the machine is still able to produce an entangled state, so that entanglement purification may still be possible. 

More explicitly,

(1) Alice possesses (or is able to generate) two or more species of quantum states. In our case, quantum states are either entangled or separable, but the case can be more general than that. 

(2) Then, Alice randomly samples a portion of the states and send them with correct labels to Bob, who then ``learns"  the states through the machine-learning method. 

(3) After that Alice sends Bob states without labels. The trained Bob will be able to classify the states as accurate as possible. \\

For the case of entanglement, one may imagine that the following hypothetical scenario:

(1) Suppose Alice and Bob are located far away, and they want to perform an experiment test, e.g. nonlocality of quantum physics. 

(2) Suppose there is a state generation machine designed to produce Bell pairs to Alice and Bob. However, the machine is faulty, in the sense that it may probabilistically generate random bits to Alice and Bob. 

(3) In this way, the overall output is exactly of the form of a Werner-like state (Eq.~(2)). An additionally assumption is that the value of p, the probability for generating the random bits are unpredictable. For example, one day, the machine may work fine with $p=0.99$; one day it may degrade to $p=0.1$. We also assume that initially, Alice and Bob do not know the pure state part in Eq.~(2). 

(4) In this scenario, suppose all Alice and Bob want to know is whether the states sent to them is entangled or not. They can achieve this goal by using the current machine-learning method, where the classifier were trained in the beginning with experimental data and correct labels, through resource consuming operations (state tomography). After that, they will just need to measure four observables only, instead of performing state tomography again.

In principle, quantum state classifiers can be constructed in a scalable way, and the involved key task is to find a suitable label. Since there is still no operationally-practical way to label the entanglement of multipartite states, classifier for entanglement remains a major challenge. Overall, the machine-learning method is beneficial when the labels require a long time to obtain e.g. by numerical methods such as quantum Monte Carlo (QMC) or density functional theory (DFT), and the numerical procedure is only required for the training set of the data only. The optimized state classifier can then be used to replace the numerical procedure to produce the labels of new data.

\subsection{The relation between CHSH inequality and entanglement}

The necessary theoretical background can be summarized as follows. Let us first introduce the standard CHSH inequality, which is applicable for any quantum state $\rho$ of a pair of qubits,
\begin{equation}
|\left\langle {ab} \right\rangle - \left\langle {a{b^\prime }} \right\rangle + \left\langle {{a^\prime }b} \right\rangle + \left\langle {{a^\prime }{b^\prime }} \right\rangle | \leqslant 2 \ ,
\end{equation}
where $<\cdot>$ represents the expectation value of observables under the measurements labeled by $a$ (or $a^\prime$) and $b$ (or $b^\prime$), for party $A$ and $B$ respectively. Violation of the CHSH inequality implies that the quantum state is entangled. However, the opposite is not true; there exists entangled quantum states which does not violate the CHSH inequality. 

In particular, let us consider the class of Werner states~$\rho_W$,
\begin{equation}
\rho_W=p\arrowvert{\psi^{+}}\rangle\langle{\psi^{+}}\arrowvert+(1-p) I/4 \ ,
\end{equation}
where $0\le p\le1$ is supposedly unknown. The density matrix can be regarded as a result of sending a Bell state, $\arrowvert{\psi^{+}}\rangle=(\arrowvert HV \rangle+\arrowvert VH \rangle)/\sqrt{2}$, through a channel mixing with a completely-mixed state~$I/4$. The Werner state is entangled when $p\ge 1/3 = 0.333$, but it violates CHSH inequality only if $p\ge 1/ \sqrt{2} = 0.707$. Therefore, without optimization, the original CHSH cannot be reliable for testing separability.

Furthermore, the violation also depends on the settings of the measurements. For example, for the following choices, $a=\sigma_z$, $a^\prime=\sigma_x$, $b=(\sigma_z+\sigma_x)/\sqrt{2}$ and $b^\prime=(\sigma_z-\sigma_x)/\sqrt{2}$, the Bell state $\arrowvert{\psi^{+}}\rangle$ violates the inequality with the maximum value $2\sqrt{2}$. However, if we change the relative phase between the two components of the state $\arrowvert{\psi^{+}}\rangle$ to $\pi/2$, even with the same settings, the Bell state no longer violates the CHSH inequality. 

\subsection{Performance of state classifier with standard CHSH inequality}
\begin{figure}[htb!]
 \centering
 \includegraphics[width=0.7\columnwidth]{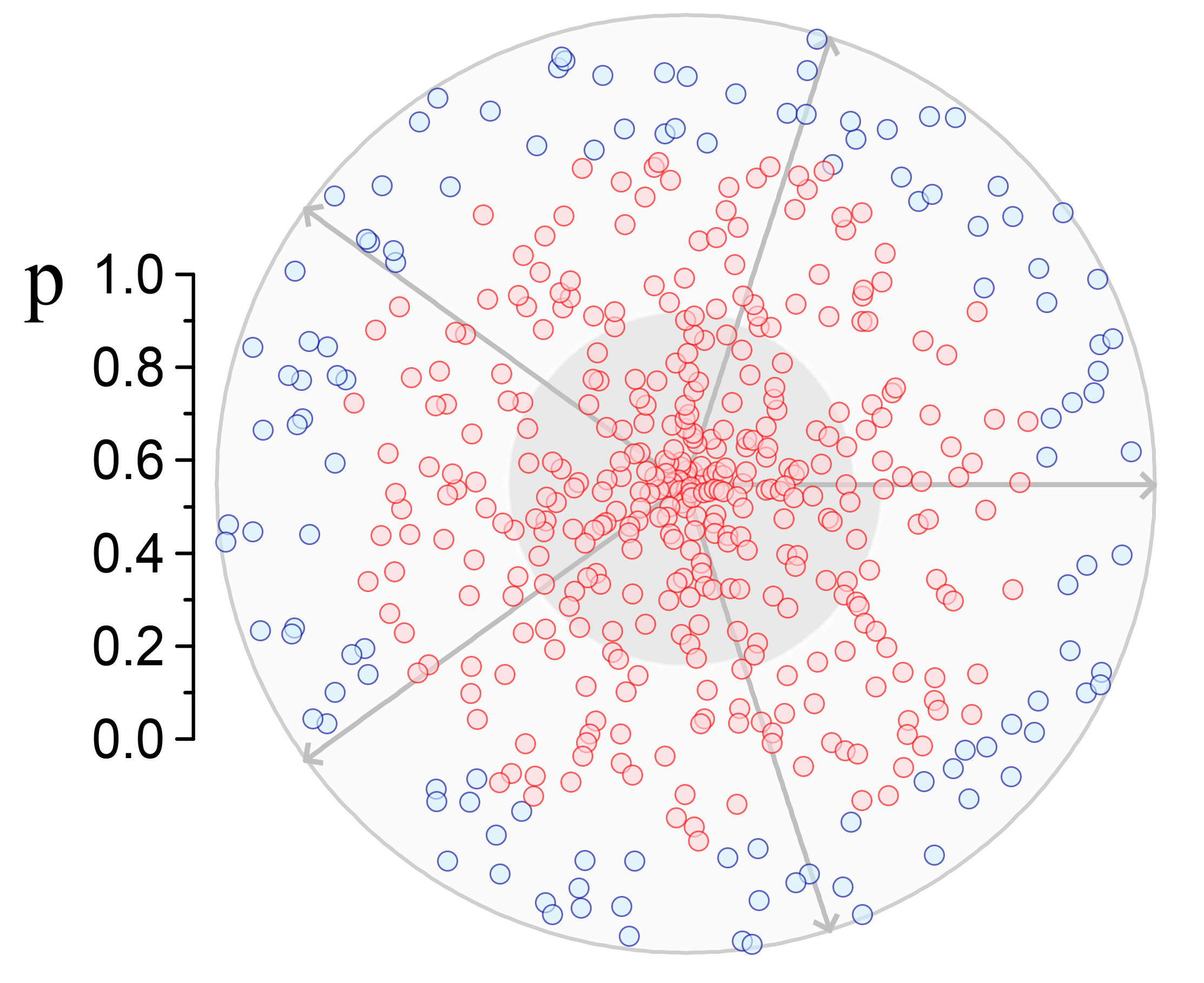}\\
 \caption{
  \textbf{Results predicted by standard CHSH inequality.} Five sections correspond to five different parameters $\theta$ which set the separable-entangled bound $p$ predicted by the PPT criteria (the light gray part marks the separable section). All the data points are labeled by standard CHSH inequality. Here blue represents the entangled label while red represents the separable label. Compared with the results shown in Fig. 3(a) in the main text, the standard CHSH inequality is apparently too loose to identify certain entangled states with $p\le0.707$.}
 \label{FIG. S1.}
\end{figure}

\section*{Acknowledgments}

The authors thank J.-W. Pan for helpful discussions. This research is supported by the National Key Research and Development Program of China (2017YFA0303700), National Natural Science Foundation of China (Grant No. 61734005, 11761141014, 11690033, 11374211), the Innovation Program of Shanghai Municipal Education Commission, Shanghai Science and Technology Development Funds, the Guangdong Innovative and Entrepreneurial Research Team Program (No.2016ZT06D348), and the Science Technology and Innovation Commission of Shenzhen Municipality (ZDSYS20170303165926217, JCYJ20170412152620376), M.-H.Y and X.-M.J. acknowledges support from the National Young 1000 Talents Plan.

\end{document}